\documentclass[12pt,preprint]{aastex}
\usepackage{amsmath}

\shorttitle{Activated asteroid P/2015 X6}
\shortauthors{Moreno et al.}

\begin{document}


\title{Dust loss from activated asteroid P/2015 X6}


\author{F. Moreno\affil{Instituto de Astrof\'\i sica de Andaluc\'\i a, CSIC,
  Glorieta de la Astronom\'\i a s/n, 18008 Granada, Spain}
\email{fernando@iaa.es}}

\author{
J. Licandro\affil{Instituto de Astrof\'\i sica de Canarias,
  c/V\'{\i}a 
L\'actea s/n, 38200 La Laguna, Tenerife, Spain, 
\and 
 Departamento de Astrof\'{\i}sica, Universidad de
  La Laguna (ULL), E-38205 La Laguna, Tenerife, Spain}}  

\author{
A. Cabrera-Lavers\affil{Instituto de Astrof\'\i sica de Canarias,
  c/V\'{\i}a 
L\'actea s/n, 38200 La Laguna, Tenerife, Spain, 
\and 
 Departamento de Astrof\'{\i}sica, Universidad de
  La Laguna (ULL), E-38205 La Laguna, Tenerife, Spain, 
\and 
GTC Project, E-38205 La Laguna, Tenerife, Spain}}

\and

\author{F.J. Pozuelos\affil{Instituto de Astrof\'\i sica de Andaluc\'\i a, CSIC,
  Glorieta de la Astronom\'\i a s/n, 18008 Granada, Spain} }


\begin{abstract}

We present observations and dust tail models 
of activated asteroid P/2015 
X6 from deep imaging data acquired at the 10.4m Gran Telescopio Canarias (GTC) 
from mid-December 2015 to late January 2016. The results of the modeling 
indicate that the asteroid has undergone a sustained 
dust loss over a two-month or longer period. The dust parameters,  
derived from multidimensional fits of
the available images, are compatible with either ice sublimation or
rotational instability processes. An impulsive event, as it could be 
associated to an impact with another body, is less likely. 
A power-law distribution of particles, with minimum and maximum radius
of 1 $\mu$m and 1 cm, and power index of --3.3 is found to be
consistent with the observations. Depending on the ejection
velocity model adopted, the particle velocities are found in the 0.3
to 10 m s$^{-1}$ range. The activation time was between 18-26 days
before discovery. The total ejected mass from that time to the most
recent observation is in the range 5-9$\times$10$^6$ kg. No dust
features giving indication of past activity earlier than the
activation time have been observed.  

\end{abstract}


\section{Introduction}
The activated asteroid P/2015 X6  
was discovered by \cite{Lilly15} from Pan-STARRS 1 images on
December 7.27, 2015. Follow-up images were obtained from several
observatories until 
December 15, 2015, by \cite{Tubbiolo15} from LPL/Spacewatch II. The
object was described as having a non-stellar coma with a narrow tail of 6 
$\arcsec$ size towards PA=75 deg. The available astrometry indicates 
that the object is asteroidal from the
dynamical point of view, as its Tisserand invariant respect to Jupiter
\citep{Kresak82} is T$_J$=3.32, derived from its orbital elements
$a$=2.756 AU, $e$=0.170, and $i$=4.56$^\circ$. To date, more than fifteen objects
of this kind have been discovered. Their activity might have been triggered by a
variety of processes, like rotational instability \citep[e.g. P/2012 F5 or P/2013 R3,][]{Drahus15,Jewitt14}, 
an impact with another object \citep[e.g. (596)
  Scheila,][]{Moreno11a}, or ice sublimation \citep[e.g. 133P or
  324P,][]{Hsieh10,Moreno11b,Hsieh15}, although 
in these latter cases no gas emission lines have been detected so far. For an in-depth review of the
activation mechanisms involved and a detailed description of the
individual objects, we refer to \cite{Jewitt15}. From
the dynamical point of view, most of those activated asteroids are
stable in time scales of 100 Myr or longer \citep{Hsieh12,Hsieh13}, so
they are 
native members of the main belt, 
although there are two objects (238P and 259P) which are unstable on
scales of only 10-20 Myr \citep{Haghighipour09,Jewitt15}.

In this paper we report observations acquired with the 10.4m GTC, and models of the dust tail
brightness of P/2015 X6, in order to characterize its
dust emission  
pattern, the activation time and the 
duration of the activity, and the total dust mass loss, and 
attempt to identify which mechanism(s) could be
playing a role in its activation.        

\section{Observations and data reduction}

Images of P/2015 X6 were acquired under photometric conditions on
the nights of 17 and 30 December 2015, and 7 and 25 January 2016. The
images were  obtained using a Sloan $r^\prime$ filter in the
Optical System for Image and Low Resolution Integrated Spectroscopy
(OSIRIS) camera-spectrograph \citep{Cepa00,Cepa10} at the GTC. OSIRIS
is equipped with two Marconi CCD detectors, each with
2048$\times$4096 pixels, providing a field of view of
7.8$\arcmin\times$7.8$\arcmin$. In order to improve the
signal-to-noise ratio  
we used 2$\times$2 binning, so that the plate scale was 0.254
$\arcsec$/px. The seeing (FWHM) in the images ranged from
  0.9\arcsec ~to 1.2\arcsec. After bias subtraction and flat-field correction a
median stack of the available frames was computed. The
resulting  
images were calibrated using standard stars, and then converted to
solar disk intensity units, which are the output of the dust tail
code. The journal of the observations is given in Table 1, where we
indicate the date of the observations (in UT and days to the asteroid
perihelion), the heliocentric ($R$) and geocentric ($\Delta$)
distances, the phase angle ($\alpha$), the position angle of the Sun
to comet radius vector (PsAng), and the angle between the Earth and
the asteroid orbital plane (PlAng). The reduced images are shown in
Figure 1, in which 
the directions to the Sun and the asteroid orbital motion are
indicated. The Earth 
crossed the orbital plane of the comet during the observational
period, being just +0.01$^\circ$ above this plane on the observation of
January 7, 2015. No dust features that could be associated to early
activity such as neck-line or trail are seen.  The Sloan r$^\prime$
magnitudes shown in Table 1 are computed on circular apertures of
radius 2$\arcsec$. To set an 
upper limit to the asteroid size, we converted the r$^\prime$
magnitudes to absolute magnitudes H$_v$ by the relation:
\begin{equation}
H_v = m_v-5\log(R\Delta)+2.5\log[\phi(\alpha)]     
\end{equation}
where $m_v$ is the apparent visual magnitude, and
$\phi(\alpha)$ is the ratio of the scattered flux at phase
angle $\alpha$ to that at $\alpha$=0$^\circ$. To obtain $m_v$ from
r$^\prime$ magnitudes we used the \cite{Fukugita96} relationship:

\begin{equation}
m_v = r^\prime -0.49(B-V)+0.11
\end{equation}

where we assumed the solar $B-V$=+0.65 \citep{Cox00}. Then, we
obtained a maximum $H_v$=+18.16. We adopt the HG formulation
\citep{Bowell89} with 
parameter $g$=0.15 (appropriate for a C-type asteroid). From the
$a$-$i$ and $a$-$e$ albedo maps of \cite{Masiero11}, a geometric
albedo of $p_v \sim$0.1 is obtained. However, all the objects found having a
sustained activity are from type C-complex
\citep{Licandro11,Licandro13} which have darker albedoes 
($p_v\sim$0.05). Then, adopting the 
\cite{Harris02} formula relating geometric albedo to asteroid
diameter $D$ (in km): 

\begin{equation}
D = \frac{1329}{\sqrt{p_v}}10^{-0.2H_v}  ,
\end{equation}

we obtain, for $p_v$ between 0.05 and 0.1, a diameter $D$ in the range 980 to 1390
m.  This provides a very stringent upper limit for
the asteroid size, as the dust
coma is likely dominating the observed
brightness in the 2$\arcsec$ aperture, but would be an useful
ingredient for the discussion of the dust models as described in Section 4.

\section{The Model}

In order to perform a theoretical interpretation of the obtained images, and to
retrieve the dust parameters, we used our Monte Carlo dust 
tail code, which has been used previously on several works on
activated asteroids and comets, including comet
67P/Churyumov-Gerasimenko, the Rosetta 
target \citep[e.g.,][]{Moreno16}. The model computes
the position on the sky plane, and their contributions to the tail
brightness, of a large amount of particles ejected
from the nucleus since a given epoch, under certain hypotheses about
the particle physical properties and size distribution, 
and the dust mass loss rate. It is assumed that 
the particles, after leaving the object's surface, are subjected
to the solar radiation pressure and gravity forces. We neglect the
object's gravity, an assumption valid for small-sized objects. Under 
those conditions, the trajectory of the particles 
becomes Keplerian, being defined by their orbital elements, which are
functions of their sizes and ejection velocities  
\citep[e.g.][]{Fulle89}. 

The  ratio of radiation
pressure to the gravity forces exerted on each particle is given by
the $\beta$ parameter,  
expressed as $\beta =C_{pr}Q_{pr}/(2\rho r)$,  where
$C_{pr}$=1.19$\times$ 10$^{-3}$ kg m$^{-2}$, $Q_{pr}$ is the radiation
pressure coefficient, and $\rho$ is the particle density. $Q_{pr}$ 
is taken as 1, as it converges to that value for
absorbing particles of radius $r \gtrsim$1 $\mu$m 
\citep[see e.g.][their Figure 5]{Moreno12a}.

A number of simplifying assumptions on the dust physical parameters 
must be made in order to make the problem tractable. Thus, the
particle density is taken as 1000 kg 
m$^{-3}$, and the geometric albedo is set to $p_v$=0.04, indicative of
dark material 
of carbonaceous composition  \citep[see e.g.][]{Moreno12a}. In
addition, a linear phase coefficient 
of 0.03 mag deg$^{-1}$ was used to correct for the phase function of
the dust particles, a typical value found for comet dust in the
1$^\circ \le \alpha \le$ 30$^\circ$ range  \citep{MeechJewitt87}. The
particles are assumed to be broadly distributed in size, with minimum
and maximum particle radii set initially to 1 $\mu$m and 1 cm, and 
following a power-law function of index $\kappa$.  In order to keep
the fitting parameters to a minimum, we set this value to
$\kappa$=--3.3, which is within the range of previous estimates of the
size distribution of particles ejected from activated asteroids and
comets. Isotropic ejection of the particles is assumed.  

The ejection velocity of the particles will depend on the activation mechanism
involved, which is unknown. In previous works of activated asteroids,
we assumed ejection 
velocities depending on the particle size following a function of the kind $v
\propto \beta^\gamma$. Values of $\gamma$ of order 0.5, derived from
simplified hydrodynamic considerations has been used
by many authors (including ourselves) 
for activity driven by ice sublimation. A weak correlation between
velocity and mass (or size) 
have been obtained for fragments ejected in collision experiments
\citep[e.g.,][]{Giblin98}, in the range 0-0.5, with mean value
of 0.23. We have obtained, in fact, 
values of $\gamma$ as low as $\sim$0.05 in the analysis of the tails
of activated asteroid (596) Scheila, that was likely the result of an
impact \citep{Moreno11a} .   

On the other hand, recent work on comet 67P, the target of Rosetta, has
shown that the velocity of the particles, measured in-situ from OSIRIS and
GIADA instruments, do not show any evident trend regarding size when
the comet was far 
from perihelion (3.6--3.4 AU) \citep{Rotundi15}.  Closer to
the Sun (3.36-2.29 AU), the size dependence, estimated with the same
instrumentation, became more apparent, being
characterized by a steep power-law function, although with a considerable
dispersion around the most probable  
value \citep[$\gamma$=0.96$\pm$0.54,][]{DellaCorte15}. Taking 
into account these results and the 
uncertainty in the particle ejection mechanism, we consider two different
models by adopting two kind of ejection velocity 
functions: a customary power-law given by $v=v_0\beta^\gamma$, which
will be called Model 1, 
and a random function of the form $v = v_1 + \zeta v_2$,
where $\zeta$ is a random number in the $[0,1]$ interval, and $v_1$
and $v_2$ are the fitting parameters. This model will be referred as
to Model 2.    

For the dust loss rate, we have adopted a half-Gaussian function whose
maximum denotes the peak dust-loss rate ($\dot{M}_0$), located
at the start 
of the particle emission event ($t_0$). The half-width at half-maximum of the
Gaussian (HWHM) is a measure of the effective time span of the 
event. 

For Model 1, we have a set of five fitting parameters 
in total, $\dot{M}_0$, $t_0$, HWHM, $v_0$, and $\gamma$, and for Model
2, we have also five fitting parameters,  namely 
$\dot{M}_0$, $t_0$, HWHM, $v_1$, and $v_2$. To fit the observed 
dust tails brightnesses, we searched for a minimum of 
the function $\chi=\sum \sigma_i$, where the summation is extended
to the four images under consideration, and     
$ \sigma_i=\sqrt{(\sum[\log(I_{obs}(i))-\log(I_{fit}(i))]^2/N(i))}$, where
$I_{obs}(i)$  and $I_{fit}(i)$ are the observed and modeled tail
brightness, and $N(i)$ is the number of pixels of image $i$. In order to avoid
regions of the images contaminated by 
field stars, the summation was restricted to pixels outside those
regions. The minimization procedure was performed by the
multidimensional downhill  
simplex algorithm \citep{Nelder65}, as 
described in \cite{Press92}. Since the algorithm 
always searches for a local minimum in the five-dimensional space of parameters,
we used different starting simplex in order to increase the probability
that the minimum of the different minima found was the deepest one. 

We performed a preliminary, zero-th order analysis, of the images by
constructing a syndyne-synchrone map for each observing date. From
those maps, we inferred that the activation time of the asteroid
should be close in time to the discovery date, owing to the absence of
dust features that could have shown-up 
at the corresponding locations of synchrones approximately two months
before discovery date or older. In particular, no neck-line or trail
features appear in the January 7th 2016 image (PlAng$\sim$0$^\circ$),
that could have 
indicated past activity. In addition, no dust condensations along the
direction of isolated synchrones, that could have indicated short
bursts of activity  
\citep[e.g., the case of P/2012 F5 (Gibbs),][]{Moreno12b}, or several
separated short bursts, as in the case of 
P/2013 P5 \citep{Jewitt13,Moreno14} are shown. According to this, it is
reasonable to start 
the search for a minimum in the function $\chi$ defined above
placing the activation time ($t_0$) between a few days before the discovery
date (102.5 days before perihelion) and about 60 days before. Regarding
the duration of the activity, the smooth variation in absolute
  magnitudes (from H$_v$=17.88 to H$_v$=18.16, see equation [1]) over the $\sim$40 days period of
observation and the 
aforementioned lack of single-synchrone dust features   
would suggest a long-lasting process and not an impulsive one,
short-duration event, like a collision with another body. In any case,  
we considered both long- and short-duration events by varying HWHM in
a wide range 
between a few days and several months in the starting simplex of the
five-dimensional parameter space search. For the peak dust mass loss,
we imposed a wide range between a minimum of 0.1 kg s$^{-1}$ and 100
kg s$^{-1}$, while for the velocities set broad limits for the parameters $v_0$,
$v_1$, and $v_2$, so that the velocities ranged from 0 to 
5$\times$10$^3$ m s$^{-1}$ (the mean velocity in the asteroid belt),
and the parameter $\gamma$ from 0.5 to 0, i.e., from typical gas drag
to nearly flat distribution of velocities.     

\section{Results and Discussion}

The resulting best-fit parameters for Models 1 and 2 are 
given in Table 2. We consider a limiting value of $\chi \le$0.15 to
consider a fit as acceptable. The uncertainties provided correspond to the range
of variation of the best-fit parameter for which $\chi \le$0.15. 
Models 1 and 2 shows very similar $\chi$ values
and very similar isophote fields, so that we only provide the results
of Model 1 in terms of those isophote fields (Figure 2, panels a1 to d1). The 
peak dust loss rate varies between 1 kg s$^{-1}$ and 1.6 kg s$^{-1}$,
while similar activation times are found for both models: 120 and 128 days
before perihelion for Model 1, and 2, respectively. The HWMH are also
close between the two models: 75 days for Model 1, and 84 days for
Model 2. The greatest difference between the models is found in the
range of ejection velocities. For Model 1, the minimum velocity,
corresponding to the largest sized particles ejected ($r$=1 cm), is 1 m
s$^{-1}$, and the maximum (for 1 $\mu$m radius particles) is only a
factor of 3 larger, because of the low value of
$\gamma$=0.12. However, for Model 2, we found a  
broader range between 0.29 and 9.8 m s$^{-1}$, giving a median value of
$\sim$5 m s$^{-1}$. If the minimum velocities
found would correspond to escape velocities, then, for Model 1, we
would obtain asteroid diameters between 1544 and 1890 m (for 
bulk densities of 3000 and 2000 kg m$^{-3}$, respectively), while for Model 2, the 
diameters would range from 450 to 550 m, for the same 
densities. According to the upper asteroid size limit imposed by the
absolute magnitude determination ($D$ in the
range 980 to 1390 m, see
above), Model 1 would not meet this constraint, and, consequently, should be
rejected.  However, there is not a priori a specific physical reason to think
that the slowest particles should be ejected at velocities near the
escape velocity, because this would depend on the ejection mechanism
involved, which is not known. 

Let us assume first a rotational instability as the cause of the
event. A short-duration event was needed to 
explain the activity in P/2012 F5 (Gibbs),   
\citep[e.g.,][]{Stevenson12,Moreno12b} and, although an 
impact with another object was invoked as the
cause, the discovery of several trailing
fragments in follow-up images by \cite{Drahus15} led them to infer a
rotational instability as the most likely mechanism.  For P/2012 F5,
we inferred quite low dust ejection velocities, of the order of 
8-10 cm s$^{-1}$ \citep{Moreno12b}. This would be in fact compatible with
rotational instability causing particles to be ejected at velocities
near the escape velocity of the parent body. In the case of asteroid
P/2013 R3,
that was observed to broke apart in several fragments, a rotationally
induced mechanism was also invoked, and the 
inferred dust speed was also low, of the order of 1 m s$^{-1}$
\citep{Jewitt14}. In contrast to P/2012 F5, however, the activity was
sustained during 2-3 month period, the fragments becoming individual
dust sources \citep{Jewitt14}. In the case of
P/2015 X6, we have also inferred a sustained activity, of the order of months,
and low velocities (2-5 m s$^{-1}$ in average, depending on the
model), so that in principle its activation mechanism would be
compatible with a rotational instability. It is important to
note, however, that we have not seen any fragments, although this is
hampered by the spatial resolution and the limiting magnitude
reachable with the used instrumentation. 

Another possibility is the occurrence of an impact that produced the excavation
of surface layers of the asteroid, exposing ices that
might exist in its interior to sunlight, or, alternatively, 
that on some surface regions 
of the asteroid the temperature becomes high 
enough when the asteroid approaches perihelion that surface ices 
can sublimate. In any of those cases, this would led to a sustained dust
activity induced by the sublimation, as is very likely the case of
activated asteroids 133P, 238P, 313P, and 324P  
\citep{Hsieh04,Hsieh11,Jewitt15,Pozuelos15,Hsieh15}, although the strongest reason for
this mechanism to occur is the fact that they show recurrent activity when 
near-perihelion. In consequence, ice sublimation would very well explain the
sustained activity we have found for this object, but more
observations during next perihelion passages would be needed 
to assess the likelihood of this mechanism. 

We could not find a solution with a comparable fit quality as the
best-fit Models 1 or 2 for an impulsive, 
short-duration event.  Thus, limiting the HWHM to 2 days or less for the
Model 1 parameters, the code converges to a
solution with a very large $\chi$=0.435. Assuming the Model 2
parameters, the results improve significantly, but still $\chi$=0.159, 
far from the best $\chi$ values of Models 1 and 2. The isophote fields from this short-burst
model (called Model 3) are shown in figure 
2, together with the results of Model 1, and the best-fit parameters
are shown in Table 2. Figure 3 display intensity scans along the tails
of the best-fit
Models 1 and 3, in which it can be seen the much better agreement of Model 1 with the data as compared
to Model 3. In summary, a short-duration event seems unlikely.

\section{Conclusions}  

From the GTC observations and the dust tail modeling of the activated
asteroid P/2015 X6, the following conclusions can be drawn:  

1) The analysis performed by a simultaneous multidimensional fit to the four
available dust tail images implies most likely a sustained
activity pattern spanning at least two months until the most recent
observation, but probably much longer. The nature of the
activation could not be determined, as both ice sublimation and  
rotational instability can be invoked as plausible mechanisms. Future
near-perihelion observations would be needed to search for recurrent
activity. The possibility of a short-duration burst of activity that
could be caused by an impact and subsequent ejection of material is
unlikely from the model fits.  

2) Two ejection velocity models, one based on a customary $v \propto
\beta^\gamma$ (Model 1) and another based on a random distribution of the
kind $v=v_1+\zeta v_2$ (where $\zeta$ is a random number in the $(0,1)$
interval) (Model 2), have been proposed. For Model 1, a nearly flat
distribution of particle velocities are found, in the range 1-3 m
s$^{-1}$.  The best-fits for Model 2 imply a wider range of
velocities in the 0.3 to 10 m s$^{-1}$ range. The quality of
the fits obtained was very similar, so that none of those models can
be favored over the other.

3) The derived dust mass loss rate is characterized by a peak value of  
1 and 1.6 kg s$^{-1}$, with a time span of $\sim$75
and $\sim$100 days, for Models 1 and 2, respectively. The total ejected dust mass since  
activation date until the most recent observation reported was 
 5-9$\times$10$^6$ kg.

4) The activation times found were of 18 and 26 days before the
discovery date, for Models 1 and 2, respectively. No dust features
that could be attributed to older activity were found.

\acknowledgments

This article is based on observations made with the Gran Telescopio
Canarias, installed in the Spanish Observatorio del Roque de los
Muchachos of the Instituto de Astrof\'\i sica de Canarias, in the island 
of La Palma. 

This work was supported by contract AYA2015-67152-R from the Spanish
Ministerio de Econom\'\i a y Competitividad. J. Licandro gratefully
acknowledges support from contract ESP2013-47816-C4-2-P.

\clearpage

\begin{figure}[ht]
\centerline{\includegraphics[scale=0.8,angle=-90]{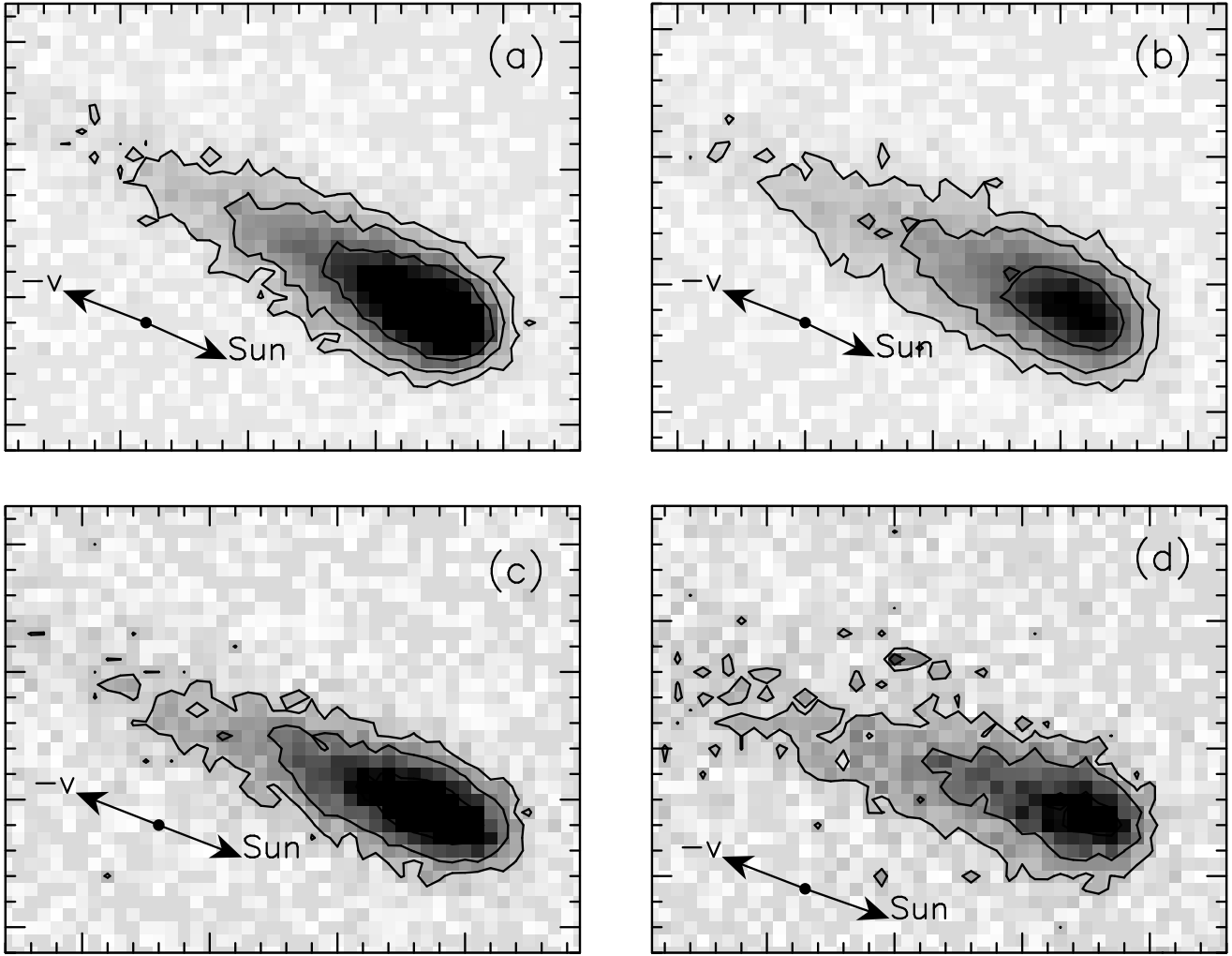}}
\caption{Median stack images of P/2015 X6 obtained with the OSIRIS
  instrument of the 10.4m GTC through a 
Sloan $r^\prime$ filter, on UT 2015 Dec 7 (a), 2015 Dec 30 (b), 2016 
Jan 07 (d), and 2016 Jan 25.  North is up, East to the left. The
directions to the Sun and the negative of the orbital velocity motion 
are shown. The spatial dimensions of the images 
at the asteroid distance (in km), 
in each panel, are as
follows: (a): 13728$\times$10677; 
(b): 14839$\times$11541; (c): 15577$\times$12115; (d): 17301$\times$13456.
   \label{fig1}}
\end{figure}

\clearpage

\begin{figure}[htb]
\centering
  \begin{tabular}{cc}
    \includegraphics[width=.40\textwidth,angle=-90]{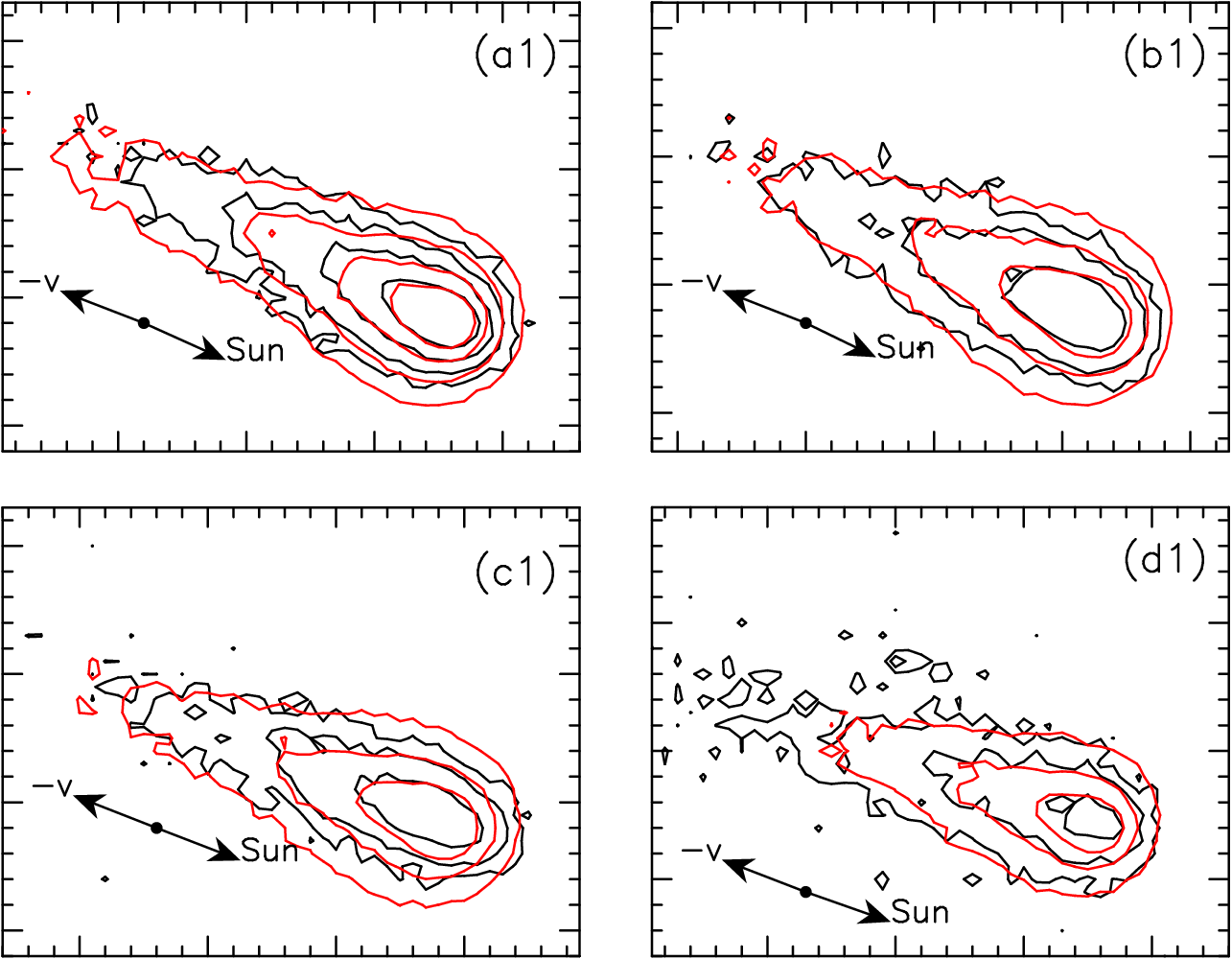} &
    \includegraphics[width=.40\textwidth,angle=-90]{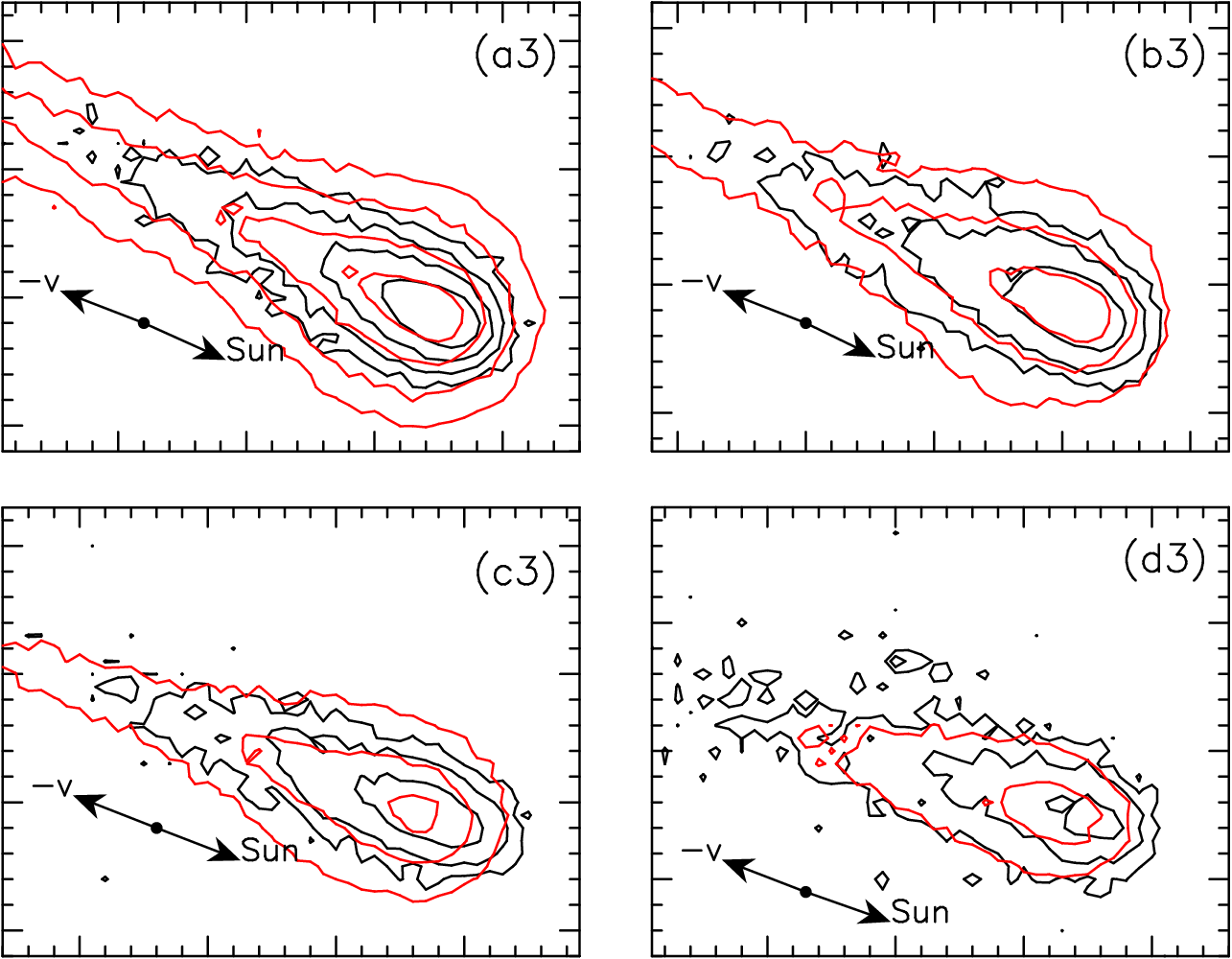} 
  \end{tabular}
  \caption{Measured (black) and modeled (red) isophotes for Model 1
    (four leftmost panels a1 to d1) and Model 3 (four rightmost panels a3 to d3). The panels
    correspond to the same dates as in figure 1. Innermost
    isophote levels are 1.5$\times$10$^{-14}$ solar disk intensity
    units, and decrease in factors of two outwards.}
  \label{fig2}
\end{figure}

\clearpage

\begin{figure}[ht]
\centerline{\includegraphics[scale=0.7,angle=-90]{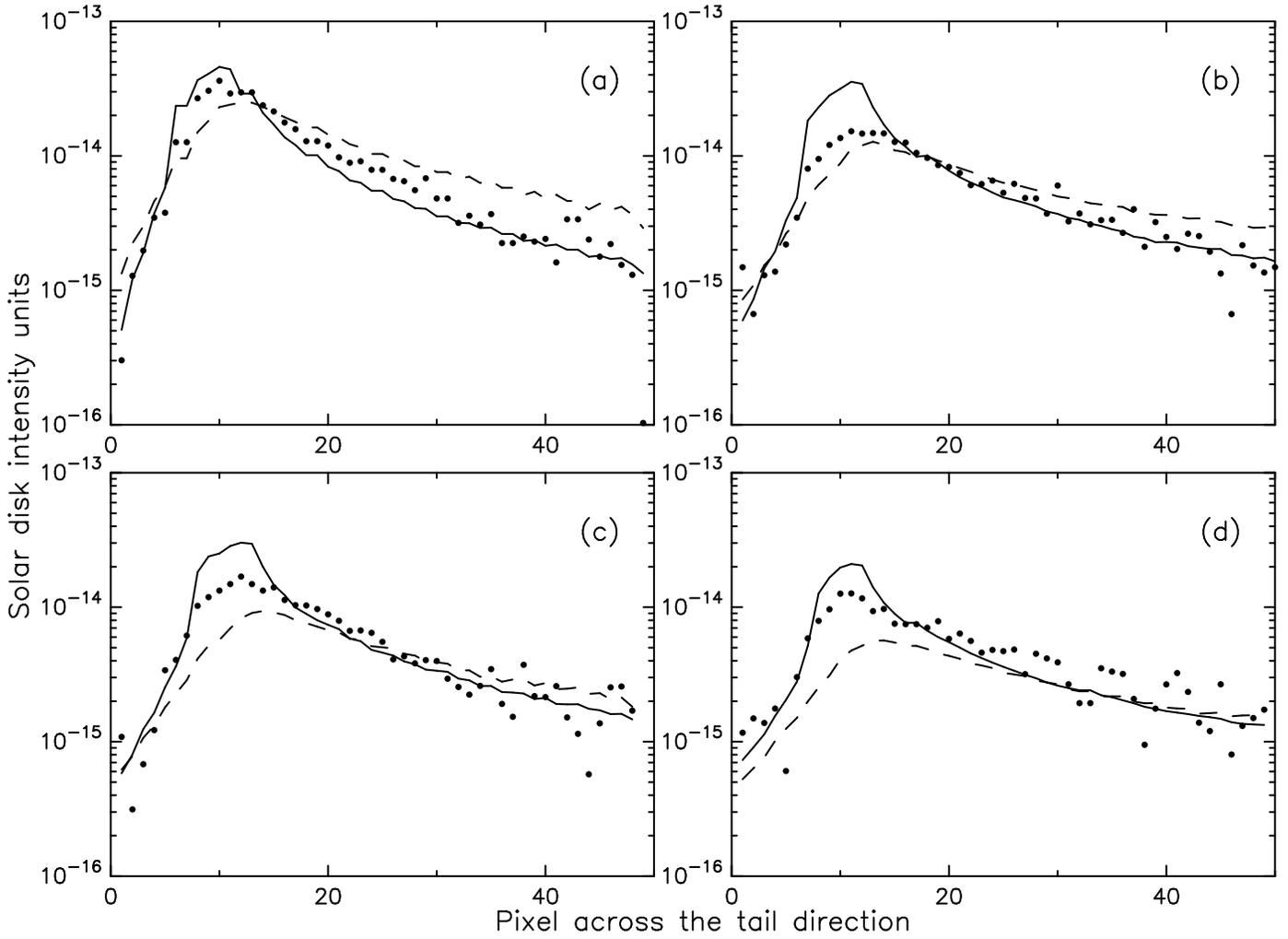}}
\caption{Scans along the direction of the tail for the observed images
  (solid circles), Model 1 images (solid lines), and Model 3 images
  (dashed lines). Labels (a) to (d) correspond to the different
  observing dates as shown in figure 1.}
\label{fig3}
\end{figure}

\clearpage

\begin{deluxetable}{cccccccc}
\tablewidth{0pt}
\tablecaption{Log of the observations}
\tablehead{
\colhead{UT} & \colhead{Days to} & \colhead{r$^\prime$} & 
\colhead{R} & \colhead{$\Delta$} &  \colhead{$\alpha$}&  \colhead{PsAng}
& \colhead{PlAng} \\
\colhead{YYYY/MM/DD HH:MM} & \colhead{perihelion} & \colhead{mag} & 
\colhead{(AU)} & \colhead{(AU)} &  \colhead{($^\circ$)}&  \colhead{($^\circ$)}
& \colhead{($^\circ$)} \\
}
\startdata
2015/12/17 22:39 & --91.8 & 21.64$\pm$0.01 & 2.327 & 1.656 & 21.13 & 65.79 &+1.00\\
2015/12/30 21:53 & --78.9 & 22.06$\pm$0.01 & 2.317 & 1.790 & 23.53 & 67.67 & +0.37\\
2016/01/07 22:55 & --70.8 & 22.28$\pm$0.01 & 2.311 & 1.879 & 24.47 & 68.65 & +0.01\\
2016/01/25 20:17 & --52.9 & 22.41$\pm$0.01 & 2.301 & 2.087 & 25.33 & 70.64 & --0.66\\
\enddata
\end{deluxetable}

\clearpage

\begin{deluxetable}{c|ccccccccc}
\tablewidth{0pt}
\tablecaption{Best-fit parameters of the models}
\tablehead{
\colhead{ } & \colhead{$\dot{M}_0$}  & \colhead{t$_0$} & 
\colhead{HWHM} &  \colhead{v$_0$} & \colhead{v$_1$} &
\colhead{v$_2$}&  \colhead{$\gamma$} &  \colhead{Total} & \colhead{$\chi$} \\ 

\colhead{ } & \colhead{(kg/s)} &\colhead{(days)} & 
\colhead{(days)} &  \colhead{(m s$^{-1}$)} & \colhead{(m s$^{-1}$)} &
\colhead{(m s$^{-1}$)}&  \colhead{ } & \colhead{mass (kg)} & \colhead{} \\
}
\startdata
Model 1 & 1.0$^{+0.3}_{-0.1}$ & --120$^{+10}_{-10}$  & 75$^{+\infty}_{-25}$ & 0.32$^{+0.09}_{-0.08}$ & -- & -- & 0.12$^{+0.05}_{-0.07}$ &  4.6$\times$10$^6$ & 0.122 \\
Model 2 & 1.6$^{+0.5}_{-0.3}$ & --128$^{+30}_{-15}$ &103$^{+\infty}_{-20}$  & -- & 0.3$^{+0.7}_{-0.3}$ & 9.1$^{+2.4}_{-4.0}$ & -- &  8.8$\times$10$^6$ & 0.120 \\
Model 3 & 41.4 & --131 &  2.0 & -- & 0.08 & 1.0 & -- & 6.0$\times$10$^6$ & 0.159 \\  
\enddata
\end{deluxetable}

\end{document}